**LETTER • OPEN ACCESS**

# Electron wave spin in a quantum well

View the article online for updates and enhancements.

## You may also like

- Optimization of the properties of $Cd_xZn_{1-x}S$ films prepared by chemical bath deposition at low $Cd^{2+}$ concentrations
  Yifan Wang and Yuming Xue

- The SAGEX Review on Scattering AmplitudesChapter 6: Ambitwistor Strings and Amplitudes from the Worldsheet
  Yvonne Geyer and Lionel J Mason

- Direct CP violation of three bodies decay process from the resonance effect
  Gang Lü, Yan-Lin Zhao, Liangchen Liu et al.

# Journal of Physics Communications

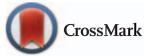

# Electron wave spin in a quantum well

Ju Gao

University of Illinois, Department of Electrical and Computer Engineering, Urbana, 61801, United States of America

E-mail: jugao2007@gmail.com

## Abstract
The particle-wave duality of the electron poses a principle question of whether the spin is a property of the particle or the wave. In this paper, the wave nature of the spin is studied for an electron inside a two-dimensional quantum well. By solving the exact $4-$spinor eigen solution to the Dirac equation, we show that a stable circulating total current density exists inside the well with a donut shaped topography. A spin value is modified by the confining geometry of the well. Our analysis also shows that a free electron Gaussian wavepacket is unstable and experiences quick decoherence.

## 1. Introduction

Spin is an essential property of an electron that has played a pivotal role in the development of quantum mechanics and quantum field theory [1]. Demonstrated by the well-known Stern–Gerlach experiment [2] exactly a century ago, spin was interpreted first by Uhlenbeck and Goudsmit [3] in 1926 as the internal angular momentum of the spinning particle electron. Since then, the use of the widely accepted $2-$spinor $\begin{pmatrix}1\\0\end{pmatrix}, \begin{pmatrix}0\\1\end{pmatrix}$ to represent the electron spin up and down [4] only solidifies the view of an abstract and internal spin of the electron. However, there are some profound obstacles of this view, including that no internal structure of the electron is observed and the electron must spin faster than the speed of light [5].

An alternative view proposed by Belinfante [6] and Ohanian [7] attributed the spin to the electron wave instead. Their works showed that the electron momentum and current densities calculated from the quantum mechanical wavefunctions have circulating flow patterns that give rise to the electron spin and magnetic moment. Here the wavefunction satisfies the Dirac equation [8]

$$\frac{1}{c}\frac{\partial}{\partial t}\psi(x) = \left(-\boldsymbol{\alpha}\cdot\boldsymbol{\nabla} - i\frac{mc}{\hbar}\gamma^0\right)\psi(x), \tag{1}$$

where $x = (t, \boldsymbol{x})$ is a four-vector and $\boldsymbol{\alpha} = \begin{pmatrix}0 & \boldsymbol{\sigma}\\ \boldsymbol{\sigma} & 0\end{pmatrix}$ are $4\times 4$ matrixes that include $2\times 2$ Pauli matrixes $\boldsymbol{\sigma}$. Any observable quantities calculated from the wavefunction thus carry the spin properties into the space permeated by the wavefunctions beyond the locality of the electron particle. The behavior of the total charge and current density expressed by the four-current $j = (\rho, \boldsymbol{j})$

$$\begin{aligned}\rho(x) &= e\bar{\psi}(x)\gamma^0\psi(x)\\ \boldsymbol{j}(x) &= ec\bar{\psi}(x)\boldsymbol{\gamma}\psi(x),\end{aligned} \tag{2}$$

calculated from the four-vector $\gamma-$matrix operator $(\gamma^0, \boldsymbol{\gamma})$, is of particular interest because it ties to the whole spectrum of electromagnetic phenomena by Maxwell equation

$$\left(\frac{1}{c^2}\frac{\partial^2}{\partial t^2} - \nabla^2\right)A = \mu_0 j, \tag{3}$$

where $A = (\phi, \boldsymbol{A})$ is the electromagnetic four-potential, providing the possibility that the wave nature of the spin can be experimentally measured and verified.

**Table 1.** Table of the constants used in the paper.

| Name and symbol | Value | Unit |
| --- | --- | --- |
| Electron charge e | $-1.6022 \times 10^{-19}$ | C |
| Electron mass m | $9.109 \times 10^{-31}$ | Kg |
| Vacuum permeability $\mu_0$ | $1.257 \times 10^{-6}$ | H m$^{-1}$ |
| Speed of light c | $2.998 \times 10^8$ | m s$^{-1}$ |
| Reduced Planck constant $\hbar$ | $1.054 \times 10^{-34}$ | JHz$^{-1}$ |

In the original paper [7], a Gaussian wavefunction is used to describe a free electron wavepacket of diameter $d$, spin up, and zero momentum expectation value

$$\psi(\boldsymbol{x}) = (\pi d^2)^{-3/4} e^{-\frac{x^2+y^2+z^2}{2d^2}} \begin{pmatrix} 1 \\ 0 \\ 0 \\ 0 \end{pmatrix}. \quad (4)$$

Plugging expression in equation (4) to equation (2) yields the Gaussian charge distribution but zero total current density $\boldsymbol{j} = 0$ without the circulating behavior. Nevertheless, the circulating flow of the current density is demonstrated through a spin current density defined by

$$j_S^k = \frac{ie\hbar}{2m} \partial_n \bar{\psi} \sigma^{kn} \psi + \frac{ie\hbar}{2mc} \frac{\partial}{\partial t} \bar{\psi} \sigma^{k0} \psi, \quad (5)$$

as part of the total current density with the help of the Gordon decomposition

$$j^k = \frac{ie\hbar}{2m}[\bar{\psi}(\partial_k \psi) - (\partial_k \bar{\psi})\psi] + \frac{ie\hbar}{2m} \partial_n \bar{\psi} \sigma^{kn} \psi + \frac{ie\hbar}{2mc} \frac{\partial}{\partial t} \bar{\psi} \sigma^{k0} \psi, \quad (6)$$

where $k$ and $n$ are indices that run from 1 to 3. The three terms in equation (6) are the convection, magnetization, and polarization current densities, respectively, where the sum of the last two terms is the spin current density defined in equation (5).

While the Gordon decomposition provides insights to the wave spin picture, the appearance of circulating flow in the spin current density but not in the total current density raises the question of whether the wave effect of the spin is cancelled internally and unobservable externally, rendering the distinction between the wave spin and particle spin ambiguous.

Settling down the wave or particle nature of the spin is intriguing not only to the better understanding of wide spectrum of physics but to the broad technical practices as well, since, if the wave nature of the spin is correct, the spin may be manipulated and utilized similarly to an optical wave for novel applications. In this paper, we continue the wave spin discussion pioneered by previous authors [5–7]. We base our analysis on rigorous solutions to the Dirac equation from which observable physical quantities are calculated. SI units are adopted for all expressions throughout the paper to facilitate practical calculations. The main physical constants used in the paper are summarized in table 1.

## 2. Dirac free electron Gaussian wavepacket

First, we shall point out that the finite-sized wavepacket of equation (4) has certain momentum values according to the uncertainty principle pd ⩾ $\hbar$, even for a wavepacket with zero momentum expectation. Just like finite Gaussian laser beam gives rise to transverse momenta that help balance the momentum conservation in a laser-electron interaction [9], the finite Gaussian wavepacket gives rise to transverse momenta that form the circulating current density as shown in the following.

The solution of the Dirac equation is a vector in the Hilbert space spanned by the eigenstates. The eigenstate for the free electron with momentum $\boldsymbol{P} = (P_x, P_y, P_z)$ is

$$\psi(x) = e^{-i\frac{1}{\hbar}Et + i\frac{1}{\hbar}\boldsymbol{P}\cdot\boldsymbol{x}} \begin{pmatrix} 1 \\ 0 \\ \frac{c}{E + mc^2} P_z \\ \frac{c}{E + mc^2}(P_x + iP_y) \end{pmatrix}, \quad (7)$$

where $E = \sqrt{m^2c^4 + P^2c^2}$ describes the relativistic energy-momentum relationship. The wavefunction that satisfies the Dirac equation should be the superposition of eigenstates of equation (7),

$$\psi(x) = N \int_{-\infty}^{\infty} \int_{-\infty}^{\infty} \int_{-\infty}^{\infty} \psi(P) u(P) dP_x dP_y dP_z$$

$$\psi(P) = e^{-i\frac{mc^2}{\hbar}t} e^{-i\frac{P^2}{2m\hbar}t} e^{-\frac{P^2 d^2}{2\hbar^2}} e^{i\frac{P_x x + P_y y + P_z z}{\hbar}}$$

$$u(P) = \begin{pmatrix} 1 \\ 0 \\ \frac{1}{2mc} P_z \\ \frac{1}{2mc}(P_x + iP_y) \end{pmatrix}, \tag{8}$$

where we keep the momentum up to the quadratic terms $P^2 = P_x^2 + P_y^2 + P_z^2$ for $P \ll mc$ when $d \gg \lambdabar_c$. $\lambdabar_c = \hbar/mc = 3.862 \times 10^{-13}$m is the reduced Compton wavelength. We shall not omit the $\frac{P}{2mc}$ factors in the spinor because they are responsible for the circulating flow of the current density, nor shall we neglect the momentum-dependent temporal factors $e^{-i\frac{P^2}{2m\hbar}t}$ because they are responsible for the decoherence of the electron wavepacket.

Equation (8) is then consolidated into a compact form after performing momentum integral over all directions:

$$\psi(x) = N e^{-imc^2 t/\hbar} G(x, y, z, t) \begin{pmatrix} 1 \\ 0 \\ \frac{1}{2mc} \frac{i\hbar}{d^2 + i\hbar t/m} z \\ \frac{1}{2mc} \frac{i\hbar}{d^2 + i\hbar t/m}(x + iy) \end{pmatrix}, \tag{9}$$

where $N$ is the normalization factor and $G(x, y, z, t)$ represents the time-dependent Gaussian profile

$$G(x, y, z, t) = \left(\frac{2\hbar^2 \pi}{d^2 + i\hbar t/m}\right)^{3/2} e^{-\frac{x^2 + y^2 + z^2}{2(d^2 + i\hbar t/m)}}. \tag{10}$$

Equation (9) is verified to be a solution of the Dirac equation, from which the total charge and current densities are calculated via the definition of equation (2)

$$\rho(x) = eN^2 G(x, y, z, t)^2 \left[1 + \frac{\lambdabar_c^2}{2} \frac{x^2 + y^2 + z^2}{d^4 + (\lambdabar_c ct)^2}\right]$$

$$\boldsymbol{j}(x) = ecN^2 G(x, y, z, t)^2$$

$$\times \frac{\lambdabar_c}{d^2 + i\lambdabar_c ct}(-y, x, \lambdabar_c ct \frac{d^2 + i\lambdabar_c ct}{d^4 + (\lambdabar_c ct)^2} z). \tag{11}$$

Equation (11) recovers the circulating flow for the total current density, where the $x$ component of $\boldsymbol{j}$ is a function of $y$ and vice versa. The above charge and current densities are time-dependent, but at $t = 0$ they reduce to the same result as the restricted wavepacket in reference [5].

As time lapses, equation (11) shows that decoherence occurs quickly for both the charge and current densities due to the time-dependent terms $e^{-i\frac{P^2}{2m\hbar}t}$ that are not included in the previous treatment [5]. Different eigenstates have different temporal phase factors to cause dephase over time. The decoherence time can be estimated from equation (11) to be in the order of $t = d^2/(\lambdabar_c c)$, which is about $8.638 \times 10^{-13}$s for a wavepacket size of $d = 10$ n$m$. The short-lived wavepacket suggests a free electron wavepacket may be constructed but can not survive long enough for practical applications.

## 3. Dirac electron inside a two-dimensional well

It is known that stable waves can be produced by confinement. We turn our attention to a simplest confinement: a two-dimensional infinite potential well. With the progress in nano and material technologies, such configuration can be actually fabricated, providing the opportunity for a real case study.

We assume the well to have zero potential inside the region of $x = -L, L$, and $y = -L, L$ and infinite potential outside. We assume there is no confinement in the z-direction and let $P_z = 0$. The waveform of the lowest energy eigenstate is found

$$\psi(\boldsymbol{x}) = Ne^{-i\frac{1}{\hbar}Et}\cos\frac{\pi x}{2L}\cos\frac{\pi y}{2L}$$

$$\times \begin{pmatrix} 1 \\ 0 \\ 0 \\ \eta i \tan\frac{\pi x}{2L} - \eta\tan\frac{\pi y}{2L} \end{pmatrix}, \tag{12}$$

where $\eta = \frac{\hbar c}{E+mc^2}\frac{\pi}{2L}$ is a dimensionless geometric factor, which becomes larger for tighter confinement. We study the ground state for this work and leave the excited states for future discussion.

The normalization factor $N = 1/(L\sqrt{1+2\eta^2})$ is obtained after invoking the unity equation $\int_{-\infty}^{\infty}\int_{-\infty}^{\infty}\psi^{\dagger}\psi dx dy = 1$.

The energetic phase factor $e^{-i\frac{1}{\hbar}Et}$ is separated entirely from the spatial wavefunction and the spinor. The separation of variables is verified by plugging equation (12) into the Dirac equation (1) to obtain the eigenvalue equation:

$$\left[-i\frac{1}{c}\frac{E}{\hbar} + i\frac{mc}{\hbar} + 2i\left(\frac{\pi}{2L}\right)^2\frac{\hbar c}{E+mc^2}\right]\psi(\boldsymbol{x}) = 0, \tag{13}$$

that gives the energy eigenvalue for the ground state,

$$E = \sqrt{m^2c^4 + 2\left(\frac{\hbar c\pi}{2L}\right)^2}. \tag{14}$$

Equation (12) is the exact ground eigenstate for the Dirac electron in the two-dimensional quantum well, which results in the time-independent charge density

$$\rho(\boldsymbol{x}) = eN^2\left(\cos\frac{\pi x}{2L}\cos\frac{\pi y}{2L}\right)^2\left[1 + \eta^2\left(\tan^2\frac{\pi x}{2L} + \tan^2\frac{\pi y}{2L}\right)\right], \tag{15}$$

and current density

$$\boldsymbol{j}(\boldsymbol{x}) = ecN^2\left(\cos\frac{\pi x}{2L}\cos\frac{\pi y}{2L}\right)^2 2\eta\left(-\tan\frac{\pi y}{2L}, \tan\frac{\pi x}{2L}, 0\right). \tag{16}$$

Here the current density shows a stable circulating wave characteristic by the $x, y$-components originated from the $4-$spinor from equation (12). The $z$ component is zero, indicating the current flows in the $x, y-$ plane. The magnitude of the total current density can be found by $j(\boldsymbol{x}) = |\boldsymbol{j}(\boldsymbol{x})|$

$$j(\boldsymbol{x}) = ecN^2\left(\cos\frac{\pi x}{2L}\cos\frac{\pi y}{2L}\right)^2 2\eta\sqrt{\tan^2\frac{\pi x}{2L} + \tan^2\frac{\pi y}{2L}}, \tag{17}$$

With the help of equations (15), (16) and (17), numerical calculation is carried out to visualize the charge and total current densities. We choose a realistic dimension $L = 10$ n$m$ for the quantum well, which gives a small but appreciable geometric factor $\eta = 3.033 \times 10^{-5}$.

First, the charge distribution in figure 1 is shown confined in the well. The density peaks at the center and distributes symmetrically within the well as expected.

Next, the vector plot of equation (16) in figure 2 confirms the circulating flow pattern. The flow is predominantly circular in the middle but gradually becomes parallel to the wall at the boundaries, showing the influence from the boundaries to the spin.

The total current density distribution of equation (17) in figure 3 displays a deep crater large enough to host the charge profile. Figures 2 and 3 illustrate the picture of the total current circulating around the charge profile at a scale determined by the quantum well, in contrast to the spinning particle picture. Also the undulation on the ridge indicates the interference from the $x$ and $y$ direction waves.

The topographies for the charge and current density distribution are illustrated in figure 4. The donut-shaped circulating current density illustrates a different picture to the particle electron spin. The donut shape topography originates from the current density as a product of the charge density and velocity distribution [5], $\boldsymbol{j}(\boldsymbol{x}) = \rho(\boldsymbol{x})\boldsymbol{v}(\boldsymbol{x})$, where the velocity distribution $\boldsymbol{v}(\boldsymbol{x})$ starts zero at the center but increases radially as

$$v(\boldsymbol{x}) = c\frac{2\eta\sqrt{\tan^2\frac{\pi x}{2L} + \tan^2\frac{\pi y}{2L}}}{1 + \eta^2\left(\tan^2\frac{\pi x}{2L} + \tan^2\frac{\pi y}{2L}\right)}, \tag{18}$$

which is bounded by the speed of light $c$, thus avoids the aforementioned superluminal obstacle in the particle spin model.

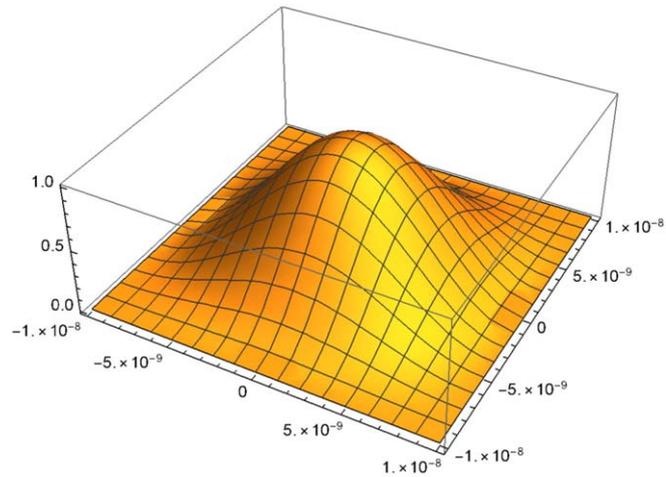

**Figure 1.** The charge density distribution plot for the ground eigenstate of an electron in the quantum well of $L = 10$ nm. The z axis represents the charge density of relative unit.

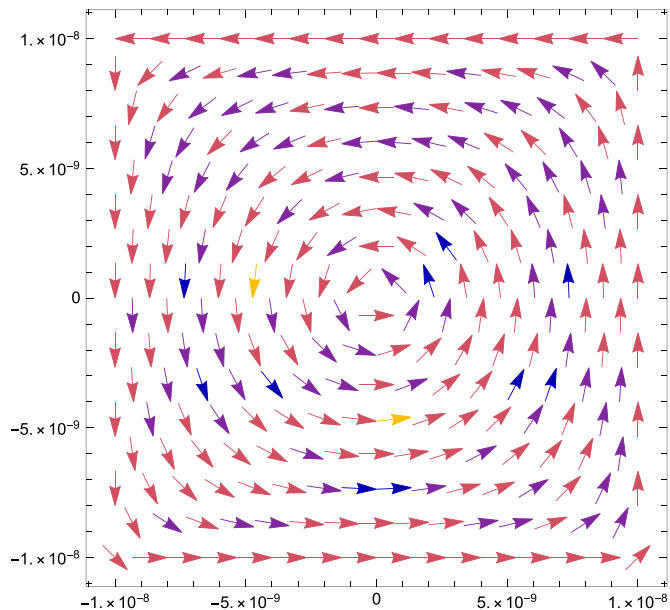

**Figure 2.** The vector plot of the total current density for the ground eigenstate of an electron in the quantum well of $L = 10$ nm.

Next we calculate the square of the spin angular momentum by using the same eigen wavefunction of equation (12) that produces the circulating total current density

$$S^2 = \int_{-\infty}^{\infty}\int_{-\infty}^{\infty} \psi^\dagger(x)\left(\frac{\hbar}{2}\boldsymbol{\Sigma}\right)^2 \psi(x)\,dxdy = \frac{3}{4}\hbar^2, \quad (19)$$

where $\boldsymbol{\Sigma} = \frac{1}{2i}\boldsymbol{\alpha} \times \boldsymbol{\alpha}$ is the Dirac spin angular momentum operator. The result is consistent with the known electron angular momentum value of $\sqrt{\frac{1}{2}(\frac{1}{2} + 1)}\,\hbar$.

The spin vector is also evaluated by the eigen wavefunction and the Dirac spin operator

$$\boldsymbol{S} = \int_{-\infty}^{\infty}\int_{-\infty}^{\infty} \psi^\dagger(x)\frac{\hbar}{2}\boldsymbol{\Sigma}\psi(x)\,dxdy = \left(0,\ 0,\ \frac{\hbar}{2}\frac{1 - 2\eta^2}{1 + 2\eta^2}\right), \quad (20)$$

which shows that the spin vector is perpendicular to the current vector but has its value $S_z = \frac{1}{2}\frac{1 - 2\eta^2}{1 + 2\eta^2}\hbar \neq \frac{1}{2}\hbar$ modified by the geometric factor $\eta$. Note that $\eta$ reduces to zero when the confinement is lifted.

The above results can be analyzed through the commutation relations between the spin operators and the Hamiltonian in equation (1), $H = -\boldsymbol{\alpha} \cdot \boldsymbol{\nabla} - i\frac{mc}{\hbar}\gamma^0$. Here $\left[H, \left(\frac{\hbar}{2}\boldsymbol{\sigma}\right)^2\right] = 0$, which means the eigenstates of the Hamiltonian are also the eigenstates of the square of the spin angular momentum, hence the spin eigenvalue is preserved. On the other hand, the z-component spin does not commute with the Hamiltonian in general, $[H, \frac{\hbar}{2}\Sigma_z] \neq 0$, unless the electron travels in z-direction, which is not the case here. This means the eigenstates of the Hamiltonian are not the eigenstates of $\frac{\hbar}{2}\Sigma_z$, hence the expectation value of the z-component spin differs from $\frac{1}{2}\hbar$.

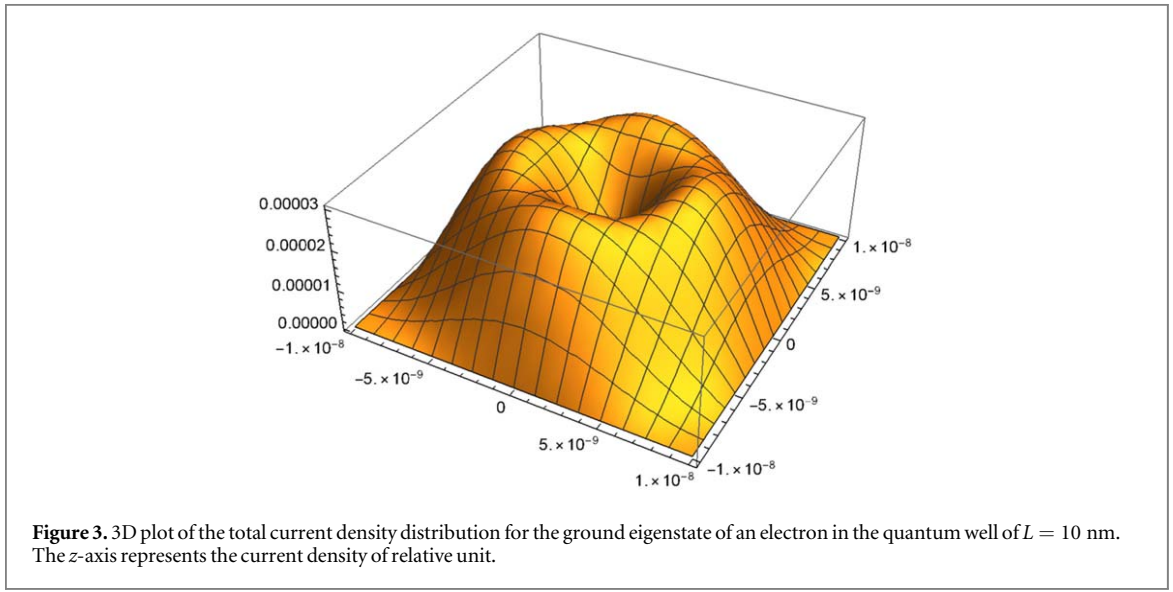

**Figure 3.** 3D plot of the total current density distribution for the ground eigenstate of an electron in the quantum well of $L = 10$ nm. The z-axis represents the current density of relative unit.

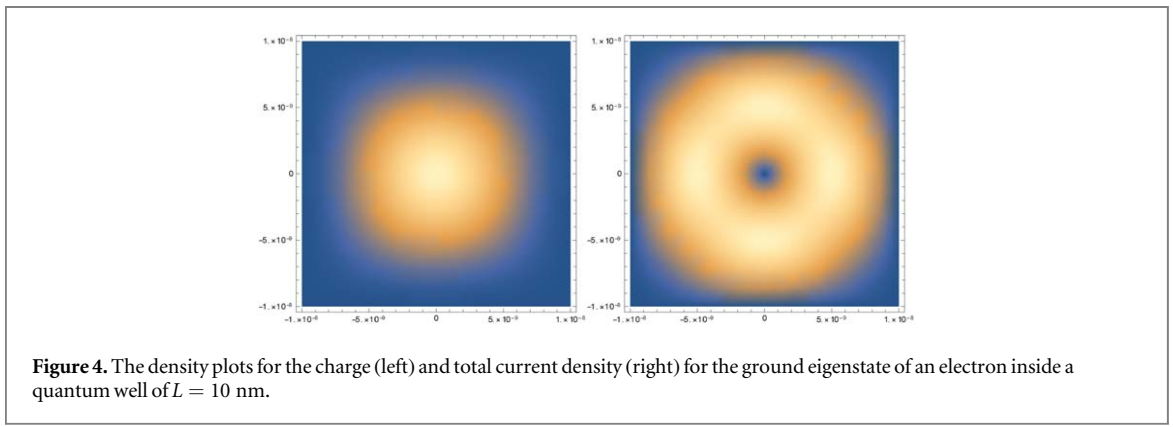

**Figure 4.** The density plots for the charge (left) and total current density (right) for the ground eigenstate of an electron inside a quantum well of $L = 10$ nm.

The above analysis highlights the theme of the paper that the spin is not a local and isolated property of the electron, but rather a wave property that can be perturbed and modified by the external conditions. We show that the impact of the confinement on the spin can be quantified as in equation (20).

## 4. Discussion and conclusion

In this work, we show that a stable circulating current exists outside the particle electron and takes up the whole space of the quantum well. The charge moves under the speed of the light at any point of the space. The above results are derived from the exact 4 − spinor eigen solution to the Dirac equation. Therefore, the spin is encoded globally in the Dirac electron wave instead of pertaining locally to the electron particle. The spin value is consequently modified by the confinement geometry as a result of its wave nature.

We also study the wave spin of a free electron by analysing the Gaussian wavepacket with the superposition of a complete set of plane-wave eigenstates of the Dirac equation. We find that the free electron wavepacket is not stable and experiences quick decoherence.

The wave nature of the spin can have critical implications for our understanding of fundamental physics and emerging technologies. It shall inevitably be involved in discussions of spin effects in solids, quantum entanglement and EPR-inspired phenomena. These topics are intended for follow-up works where the spin is no longer treated as a local property of the electron.

On the technological side, the electron spin can potentially be used as a building block for quantum computing [10] and quantum information [11]. As shown in the paper, isolation of the wave spin from the environment is impossible due to the boundary effects on the wave. Any geometric and material changes to the boundaries can perturb the spin, which leads to errors in the storing, transmitting, and processing of information. Novel schemes can be devised for spin readout by probing and even mapping the total current density that is of comparable dimensions to modern semiconductor and nano devices, opening up fresh path to holographic and parallel quantum computing and information processing.

## Acknowledgments

The author would like to thank F. Shen, D. S. Guo and Y. S. Wu for the constructive and stimulating discussions. The author would also like to thank J. Y. Gao for the careful proof-reading of the manuscript and intriguing thoughts of the wave spin implication for molecular biology. The author also wants to thank referees for their helpful and insightful comments. This work is partially funded by Suzhou Leijing Electronics Tech. Co., Ltd.

## Data availability statement

All data that support the findings of this study are included within the article (and any supplementary files).

## ORCID iDs

Ju Gao  https://orcid.org/0000-0001-5261-0896